\documentclass[conference,a4paper,twocolumn]{IEEEtran}
\usepackage{booktabs} 
\usepackage{graphicx,multicol,latexsym,amssymb,subcaption,graphicx}
\def\bM{{\bf M}}

\usepackage[utf8]{inputenc} 
\usepackage[T1]{fontenc}
\usepackage{url}
\usepackage{ifthen}
\usepackage{cite}
\usepackage[cmex10]{amsmath} 

\begin{document}
\title{Blockchain moderated by empty blocks to reduce the energetic impact of crypto-moneys}


\author{%
  \IEEEauthorblockN{Philippe Jacquet}
  \IEEEauthorblockA{Nokia Bell Labs\\
                    Nozay, France\\
                    Email: philippe.jacquet@nokia-bell-labs.com}
  \and
  \IEEEauthorblockN{Bernard Mans}
  \IEEEauthorblockA{Macquarie University\\
                    Sydney, Australia\\
                    Email: bernard.mans@mq.edu.au}
}


\maketitle

\begin{abstract}
While cryptocurrencies and blockchain applications continue to gain popularity, their energy cost is evidently becoming unsustainable. In most instances, the main cost comes from the required amount of energy for the Proof-of-Work, and this cost is inherent to the design. 
In addition, useless costs from discarded work (e.g., the so-called Forks) and lack of scalability (in number of users and in rapid transactions) limit their practical effectiveness.

In this paper, we present an innovative scheme which eliminates the nonce and thus the burden of the Proof-of-Work which is the main cause of the energy waste in cryptocurrencies such as Bitcoin. We prove that our scheme  guarantees a tunable and bounded average number of simultaneous mining whatever the size of the population in competition, thus by making the use of nonce-based techniques unnecessary, achieves scalability without the cost of consuming a large volume of energy. The technique used in the proof of our scheme is based on the analogy of the analysis of a green leader election. 
The additional difference with Proof-of-Work schemes (beyond the suppression of the nonce field that is triggering most of the waste), is the introduction of (what we denote as) ``empty blocks'' which aim are to call regular blocks following a staircase set of values. Our scheme reduces the risk of Forks and provides tunable scalability for the number of users and the speed of block generation.
We also prove using game theoretical analysis that our scheme is resilient to unfair competitive investments (e.g., "51 percent" attack) and block nursing. 
\end{abstract}


\newtheorem{theorem}{Theorem}
\newtheorem{lemma}{Lemma}

\section{Introduction}

Popular cryptocurrencies have now been effectively in use for more than ten years ({\it e.g.} Bitcoin~\cite{bitcoin}). In addition to the increasing appetite for its key financial aspects (e.g., fast, transparent, secure and private), their potential to facilitate, verify, or enforce the negotiation or performance of various transactions provides an important innovative service that will have a disruptive impact for many applications.

Actual cryptocurrencies are completely decentralized protocols. They are based on the blockchain concept where transactions are regrouped by blocks and committed in a common distributed directory connected in peer to peer organization. Its novelty comes from the combination of existing concepts and results from distributed computing, cryptography and game theory. The transactions and the blocks are chained via a chain of hash function which is too expensive to forge and falsify. The blockchain allows that each block and each transaction can be checked back by anyone as if there would be in a gigantic distributed accounting book. The block has also the function of tracing and controling the inflation of the Bitcoin volume and facial value. This is done by giving a reward to the block miner (the action of proposing a block) which is accepted by distributed consensus when the block is ``committed'' (e.g., in Bitcoin, followed by six newer blocks). 
A block within a forked branch that is not committed disappears.

One drawback of  cryptocurrencies such as Bitcoin is the moderation of block mining rate. With mining reward as a tool in the control of volume of values, there is a fierce competition in block mining. Some individuals may be tempted to mine multiple blocks in order to augment their revenue or bend the system in their favor (e.g., by double spending). 

In order to avoid that too many simultaneous blocks are mined and proposed for commitment, Bitcoin requires a “proof of work” by forcing the block miner to execute a large number of CPU cycles via iterated hash computations before submitting his/her block. 
The idea of computational puzzle as proof of work is not new, it appears first in the seminal work of Dwork and Noar~\cite{dwork92} and was later suggested to avoid Sybil attack by Aspnes et al.~\cite{aspnes06}. The proof of work provides scarcity and uniqueness that help reduce the individual block mining rate.

However the “Proof of Work” (PoW) presently costs 100,000,000,000 CPU cycles (around 10 minutes on a current PC) and if one million miners contend at the same time, then the energy cost of mining one Bitcoin exceeds the value of the Bitcoin itself. Furthermore this cost is expected to quadratically grow or at least linearly grow with the number of contending miners leading to a protocol whose energy cost can absolutely not be sustainable in the near future. 

In order to avoid these catastrophic consequences, a plethora of alternatives have been suggested and analysed  (e.g.~\cite{garay15,garay17}), including variants of proofs of stake (e.g.,~\cite{ethereum,Activity2014}) or various proofs of ``something'' (e.g., proof of Exercise~\cite{Shoker2017}). The later are difficult to implement since theydemand external element of proof (fortune, volume and age of coins, etc) and are difficult to extrapolate to blockchain system not related to currencies. Other attempts to tackle the huge waste generated by the Proof of Work as implemented in Bitcoin have been aiming at turning them into work that is actually useful: e.g., solving practical problems have been investigated in~\cite{Useful2017}, extended to present PoWs that are based on the Orthogonal Vectors, 3SUM, and All-Pairs Shortest Path problems in~\cite{Ball2017,Ball2018}, and can be used to allow for much quicker verification and zero-knowledge PoWs. Yet it is unclear how this creates incentives for large amount of transactions, and how it guarantees Liveness (and in particular freedom from starvation if no one volunteers to solve a practical problem). Also it still remains paramount to reach agreement and to prevent double spending while providing a way to mint the currency. 

In this paper, we focus on the unsustainable energy cost and propose a cryptocurrency protocol moderated through a "green" mining protocol in such a way that it will remove the prohibitive energy cost of the proof of work while keeping the blockchain mining input rate between reasonable limits. Those limits can be arbitrarily tuned {\it a priori}, while the proof of work CPU parameter must be constantly updated. In fact the system we propose can be applied to blockchain systems not related to cryptocurrencies and which necessitate high frequency updates~\cite{mirror}.

Our validation of this scheme is inspired by ``green'' leader elections~\cite{green},~\cite{cichon}. At its core our model implements a scheme which guarantees a tunable and bounded average number of simultaneous mining whatever the size of the population in competition, like a Leader Election. In addition to public verifiability, our scheme provides liveness and fairness: the unique chain grows and the probability of electing each party is independent to its relative computational power. 
Our model is “permissionless”: there is no explicit membership protocol, anyone can join the system. Our model also offers the safety property (blocks are eventually committed and cannot be removed). Finally our scheme reduces the risk of Forks (divergent chains of blocks that need to be eventually discarded) and provides tunable scalability for the number of users and the speed of block generation. Interestingly our scheme could be adapted to the aspirational "Proof of Kernel"~\cite{kernel} which consists into reducing the population allowed to contend for mining. In this case the PoW is maintained but with a much reduced difficulty just in order to prevent block forgery.

The exact contribution of our work and the structure of the paper are as follows:
\begin{itemize}
\item We introduce our green blockchain protocol in Section~\ref{sect:green}. Our protocol uses a block  generation scheme that does not use a nonce in its Proof of Work. The novelty of our protocol comes from a call field that regulates the rate of creation of transaction blocks, in addition to empty blocks.
\item We prove the performance analysis of our protocol in Section~\ref{sect:analysis}. The methodology of the proof follows a distributed leader election via collision scheme (that has an interest on its own). The technique garantees a tunable and bounded average number of simultaneous mining, whatever the size of the population in competition for block generation, thus offering full scalability.
\item In section~\ref{sec:distributed} we present (in two steps) a  protocol for implicit empty blocks that renders the scheme completely distributed, tunable and scalable. As the scheme also reduces the chances of forks, it further reduces the energy waste. We provide the performance analysis of the implicit empty block scheme.
\item In Section~\ref{blocknursing}, we prove the resilience of our scheme: as energy wasting PoW farming is no longer possible with our new scheme, we prove that block mining predators that create attack by (what we call) "block nursing", have such a low expectation of success that it renders attacks, such as "51 percent" attack, prohibitive and unworthy. 
\item We give concluding remarks and highlight future works in Section~\ref{conclusion}.
\end{itemize}

\section{The Green Coin mining protocol}
\label{sect:green}
\subsection{Parameters and format}
Our \emph{Green Coin protocol} is based on the following parameters:
\begin{itemize}
\item an integer $k$;
\item a vector of increasing probabilities of length $k+1$: $P_0,P_1,\ldots,P_k$, with $P_k=1$;
\item an upper bound $N$ on the maximum admissible number of contending blocks, that can be set as large as possible.
\end{itemize}
Typical values for these parameters would be $k=8$, $N=2^{32}$ are examples. The green coin leader protocol consists in authorizing the mining of an average number of order $N^{1/k}$ out of $n\le N$ contending blocks. 

As per Bitcoin, each block of the Green Coin protocol will carry some values obtained by hash function. We denote $h$ the length of the hash value fields in the block. A typical value is $h=256$ (32 bytes). 
But while Bitcoin has only one kind of block, Green Coin relies on two kinds of blocks:
\begin{itemize}
\item the {\it regular} block containing transactions;
\item the {\it empty} block which does not contain any transactions.
\end{itemize}

The regular block format is similar to Bitcoin block format but with the difference that it does not contain a {\it nonce} field but instead a {\it call} field. 
The call field dictates the required criteria for the next hash value field. 
The following table gives an abstract format listing the most significant fields:

\begin{center}
\begin{tabular}{|c|l|}
\hline
1&hash of previous block\\
\hline
2&date\\
\hline
3&list of transactions\\
...& ...\\
\hline
4&call value\\
\hline
5&hash of the block\\
\hline
\end{tabular}
\end{center}

The empty block looks similar but with no transaction list. 

\begin{center}
\begin{tabular}{|c|l|}
\hline
1&hash of previous block\\
\hline
2&date\\
\hline
3&call value\\
\hline
4&hash of the block\\
\hline
\end{tabular}
\end{center}

\subsection{Mining protocol}
Initially, to simplify our presentation, we will assume that empty slots are mined by a central entity and on a restricted block server. We will discuss later how the protocol can be made distributed or how to face forgery. 

A block can be mined only if its hash value is smaller than the call value of the previous block. The call field of a regular block has always the same value which is $\lfloor 2^{h+1}P_0 \rfloor - 1$. If no regular block is mined, after a certain lapse of time the central entity mines an empty block. Unlike regular block hash values, an empty block hash value does not need to be smaller than the call value of the previous block. The call value of the first empty block should have value $\lfloor 2^{h+1}P_1\rfloor - 1$. 
As long as no new regular block is mined, the central entity mines empty block with successive call values $\lfloor 2^{h+1}P_2\rfloor - 1,\ldots,\lfloor 2^{h+1}P_k\rfloor - 1$. After $k$ mined empty blocks the call values sequence restarts. 

There are {\bf three facts} that are worth noting: 
\begin{itemize}
    \item as there is no nonce field in the regular block there is no proof of work computation required, 
    \item instead, the blocks are authorized via the call values which have the effect of regulating the mining rate, 
    \item the non-authorized blocks do not need to be submitted and this limits also the traffic generated by the block mining. 
\end{itemize}

\begin{figure}[h]
\includegraphics[width=11cm]{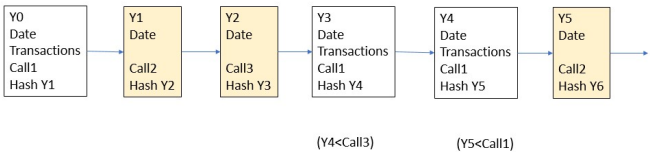}
\vskip-12.2cm
\caption{A sequence of blocks, empty blocks are grayed/yellow.}
\label{fig0}\end{figure}

{\bf Liveness.} If a call fails, {\it i.e.} no regular block is mined, then a new call will be made via a new empty block. Since $P_k=1$, the last call value is the max-value $2^{h+1}-1$. Therefore any competition of block mining will result in at least one block mined. 
Figure~\ref{fig0} displays an example of block sequences in the blockchain. The chain is enforced by the hash values $Y_0,Y_1,\ldots$ 
The value of a regular block hash should always be smaller than the call value of the previous block, regardless it is empty or regular. Empty blocks do not need to follow this rule.

\section{Performance analysis using Leader election}
\label{sect:analysis}
\subsection{Analogy of a classic leader election via collision}
Let us introduce a game of leader election to analyse the performance of our protocol and its scalability. The goal of leader election (e.g.,~\cite{Grabner,Prodinger}) is to select one among $n > 0$ players, by proceeding through a number of rounds. Note that using rounds is also commonly accepted that the Blockchain model incorporates some degree of synchrony assumptions: in a network with unbounded delay, the adversary can boost its  power just by making the non-faulty parties incur higher delays (e.g.,~\cite{decker13}). We will show that these assumptions are actually unnecessary for our complete scheme in Section~\ref{sec:distributed}.

Informally, the rounds of our leader election are made via coin tossing. We assume that each competitor has a coin whose head probability is $p$ and tail probability $q=1-p$. We want to proceed to a leader election among $n\ge 1$ competitors. At the beginning all the $n$ competitors transmit. If this results into a collision then each competitor tosses the coin, only those who get a head contend for the second round, we call them the survivors. If the second round results in a collision, then a second coin tossing occurs and a subset of survivors contend, and the protocol continues until none survive. In this case we take as result the last non empty set of survivors. 

Usually there is an additional procedure in order to reduce the survivor population to a single leader using a collision algorithm. In our case, this is not necessary as long as the subset of survivor is of reasonable size:  
we will prove using analytic combinatorics that the average size of the last non empty survivor set is $\frac{q}{p\log(1/p)}$ with some fluctuations of small amplitude. Note that this value is finite and independent of the initial number contenders (although small fluctuations may enlight some non trivial dependencies). This number can also be tuned to be small or large by tuning the parameter $p$. This can be done by compromising with the average number of rounds which tends to $\log_{1/p}n$ when $n\to\infty$. 



If we wanted to imitate this process for the call values in the empty block, we will need a sequence $(P_0,P_1,\ldots)$ such that $P_\ell=p^\ell$ represents an exponential descending staircase but this will create the case that after the first call one would have already $n$ mined blocks. 

However we can use our protocol to create the ascending exponential staircase effect when there is a large yet fixed limit $N$ to the number of simultaneous contenders.
In this case we suppose that $P_0=\frac{1}{N}$ and that $P_\ell=\frac{1}{N}p^{-\ell}$ for $\ell\le k$. In order to have $P_k=1$ one must have $p=\frac{1}{N^{1/k}}$. If $N=2^{32}$ and $k=8$ we will have $p=\frac{1}{16}$ and the call sequence will be:

{\small
\begin{center}
\begin{tabular}{|c|c|c|}
\hline
call value rank & Probability sequence & call value \\
$0 \leq \ell \leq k=8$ & $P_\ell = N^{\frac{\ell}{k}-1}$ & $\lfloor 2^{h+1}.P_\ell \rfloor -1$ \\
 \hline
0 (initial)& $2^{-32}$ & $2^{224}-1$\\
1& $2^{-28}$ &$2^{228}-1$\\
2& $2^{-24}$ &$2^{232}-1$\\
3& $2^{-20}$ &$2^{236}-1$\\
4& $2^{-16}$ &$2^{240}-1$\\
5& $2^{-12}$ &$2^{244}-1$\\
6& $2^{-8}$ &$2^{248}-1$\\
7& $2^{-4}$ &$2^{232}-1$\\
8& 1 &$2^{256}-1$\\
\hline
\end{tabular}
\end{center}
}

We denote $\bM_n$ the average number of regular mined block when $n$ blocks are in competition after a regular block. Note that an important yet regular assumption in this paper is that successive calls of a hashing function are independent. Of course {\it stricto sensu} this is not true since hash value determinations are deterministic computations, but traditionally the better successive hash values imitates independent random variables, the better is a hash function. In fact this argument is the foundation for accepting the resilience of blockchains (made of successive hash computations) against attacks.

Let $\bM_n^\ell$ denote the number of blocks mined after the $\ell$th empty block under the condition that all the previous empty blocks resulted into no regular block mined. We have $\bM_n=\bM_n^0$ and since $P_k=1$: $\bM_n^k=n$. Since the fact that the successive calls to hash values are assumed independent, the number of blocks called by the $\ell$th empty block is a binomial random variable $B(n,P_\ell)$ of probability $P_\ell$. For $\ell<k$ when the binomial variable is $B(n,P_\ell)=0$ which occurs with probability $(1-P_\ell)^n$, no block is mined after the $\ell$th block and $\bM_n^\ell=\bM_n^{\ell-1}$. Thus for all integers $m$, $m>0$:
\begin{eqnarray}
P(\bM_n^\ell=m)&=&\binom{n}{m}P_\ell^{m}(1-P_\ell)^{n-m}\nonumber\\
&&+(1-P_\ell)^nP(\bM_n^{\ell+1}=m)
\label{eqMnu}\end{eqnarray}
and
\begin{equation}
P(\bM_n^\ell=0)=\delta(n).
\end{equation}
where $\delta(n)=1$ if $n=0$ and zero otherwise, also known as Kroenecker symbol.

Let $M^\ell_n(u)=E[u^{\bM_n^\ell}]$.
\begin{lemma}
We have for $\ell<k$
\begin{eqnarray*}
M_n^\ell(u)&=&\left(1+P_\ell(u-1)\right)^n-(1-P_\ell)^n\\
&&+(1-P_\ell)^nM_n^{\ell+1}(u)
\end{eqnarray*}
and $M_n^k(u)=u^n$.
\label{lemMnu}\end{lemma}
\begin{IEEEproof}
This is a direct application of Equation~(\ref{eqMnu}).
\end{IEEEproof}
\begin{lemma}
We have the identity
\begin{equation}
E[\bM_n]=nP_0+\sum_{\ell=1}^{\ell=k}nP_\ell\prod_{j<\ell}(1-P_j)^n
\label{eq-Mn1}\end{equation}
\end{lemma}
\begin{IEEEproof}
This is a direct application of previous lemma by using $E[u^{\bM_n}]=E[u^{\bM^0_n}]$ and $E[u^{\bM_n^\ell}]=\frac{\partial}{\partial u}M_n^\ell(1)$.
\end{IEEEproof}
\begin{theorem}
For all $n\le N$, we have the estimate $$E[\bM_n]=O(N^{1/k}).$$
\label{theo1}\end{theorem}
\paragraph*{Remark} the optimal value is $k=O(\log N)$ but lower values of integer $k$ are already interesting. 
\begin{IEEEproof}
From Equation~\ref{eq-Mn1} we get the inequalities
\begin{eqnarray*}
E[\bM_n]&\le& nP_0+\sum_{\ell=1}^knP_\ell(1-P_{\ell-1})^n\\
&\le& nP_0+\sum_{1\le\ell\le k}nP_\ell\exp\left(-P_{\ell-1}n\right).
\end{eqnarray*}
The quantity $\sum_{1\le\ell\le k}nP_\ell\exp(-P_{\ell-1}n)$ is equal to 
$$\sum_{1\le\ell\le k}\frac{n}{N}p^\ell\exp(-p^{\ell-1}n/N).$$ 
This quantity is smaller than $f(n/N)$ with $f(x)=\frac{1}{p}\sum_{\ell\in\mathbb{Z}}g(p^{\ell}x)$ with $g(x)=xe^{-x}$. The function $f(e^x)$ is periodic of period $\log p$ and therefore is bounded. 

Since $p=N^{-1/k}$, $n/N\le 1$ and $n\le N$, we get $$E[\bM_n]=O(N^{1/k}).$$
\end{IEEEproof}

\begin{figure}
\includegraphics[width=8cm]{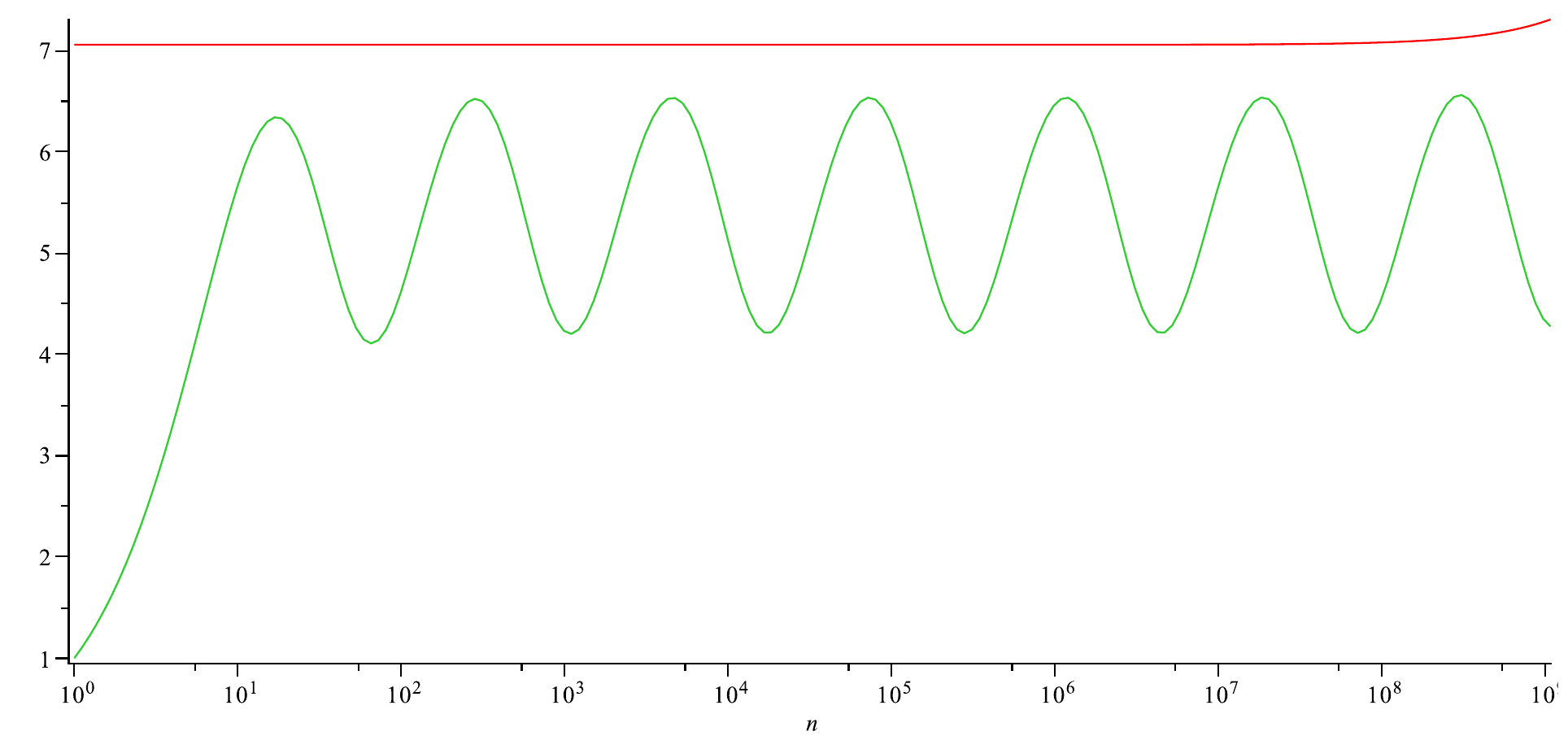}
\includegraphics[width=8cm]{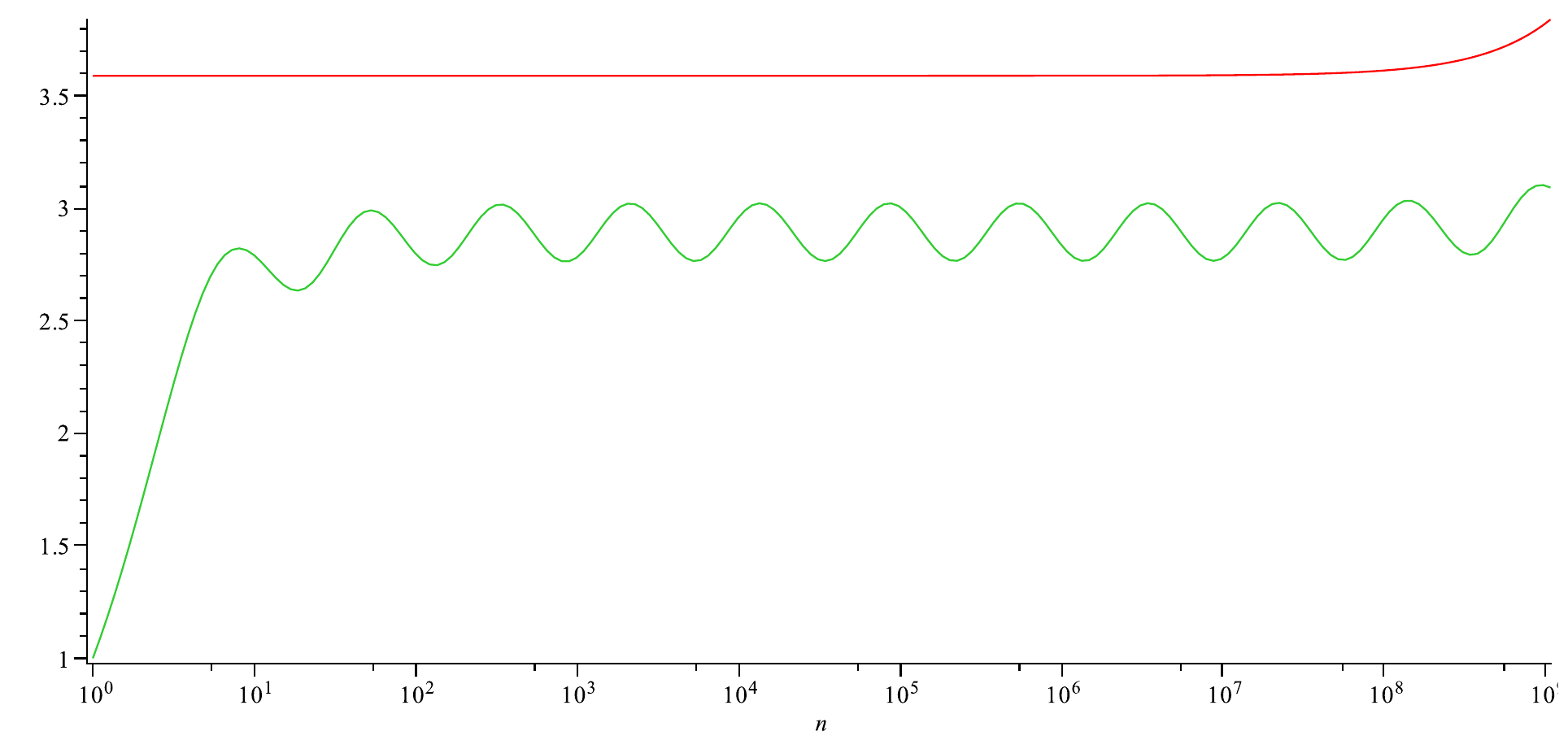}
\includegraphics[width=8cm]{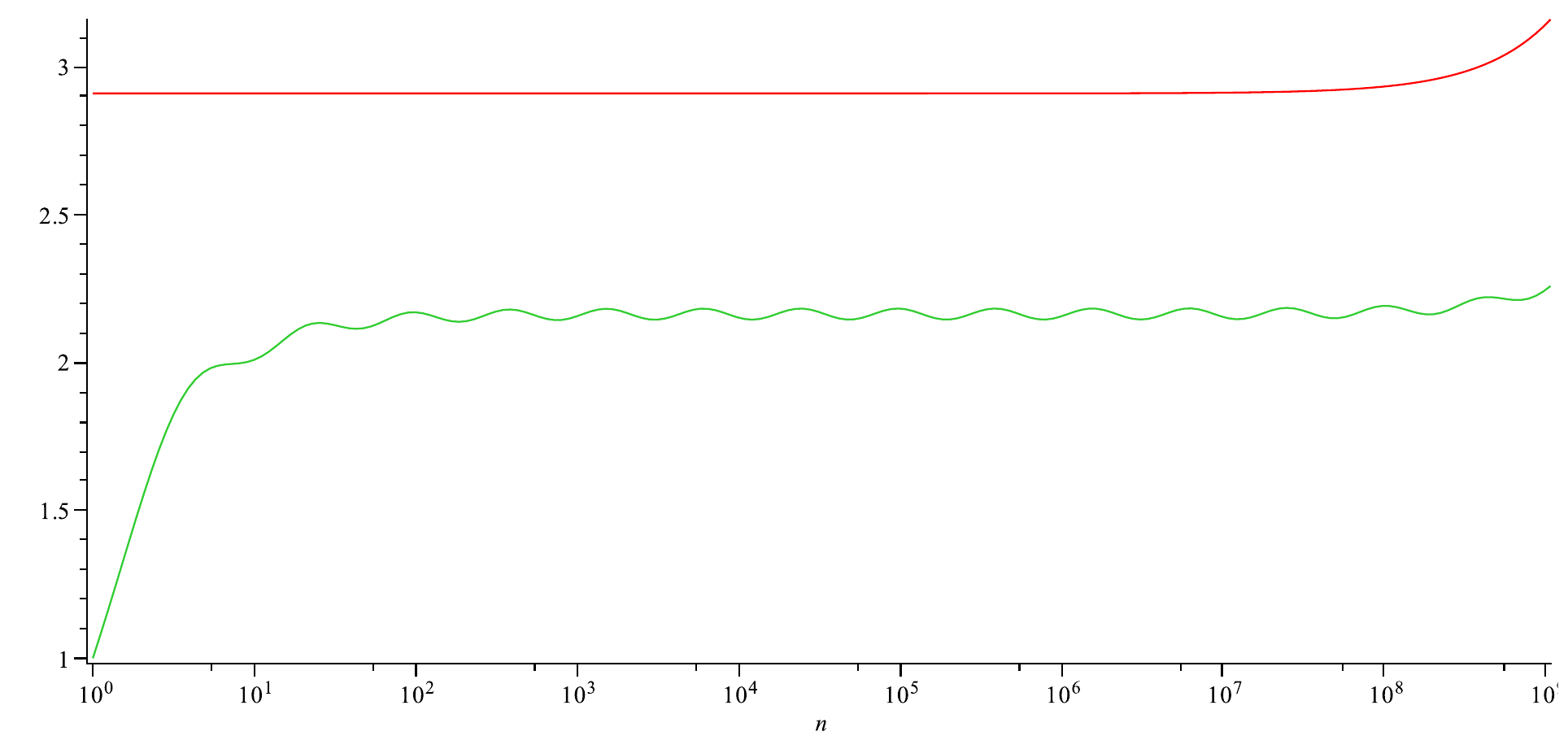}
\caption{The quantity $E[\bM_n]$ and the upper bound given by~(\ref{eq-Mn}) in red versus $n$ for $N=2^{32}$, (top) for $k=8$, (middle) $k=12$, (bottom) $k=16$.}
\label{fig1}
\end{figure}

\begin{theorem}
We have the more precise estimate
\begin{equation}
E[\bM_n]\le \frac{n}{N}+A N^{1/k}.
\label{eq-Mn}\end{equation}
with $A=\frac{1}{\log(1/p)}\sum_{k\in\mathbb{Z}}\left|\Gamma\left(1+2ik\pi/\log p\right)\right|$, $p$ denoting $N^{-1/k}$ and where $\Gamma(.)$ is the Euler "Gamma" function.
\end{theorem}
\begin{IEEEproof}
We can give an accurate estimate of the maximum value of function $f(x)$. The function $f(e^x)$ is periodic of period $\log p$ and therefore has a Fourier decomposition: 
\begin{equation}
f(e^x)=\sum_{k\in}g_k e^{2ik\pi x/\log p}
\end{equation}
where $g_k$ is the Fourier coefficient.

We have 
\begin{eqnarray*}
g_k&=&\int_{\log p}^0f(e^x)e^{2ik\pi/\log p}dx\\
&=&\frac{1}{\log(1/p)}\int_0^\infty g(x)^{2ik\pi/\log p-1}dx\\
&=&\frac{1}{\log(1/p)}\Gamma\left(1+2ik\pi/\log p\right)
\end{eqnarray*}
and then use the upper bound $|f(e^x)|\le\sum_{k\in\mathbb{Z}}|g_k|$.

In passing we get the mean value of $f(e^x)$ to be equal to $g_0=\frac{1}{\log(1/p)}$.
\end{IEEEproof}
Figure~\ref{fig1} shows the quantity $E[\bM_n]$ versus $n$ for various parameters. Notice that the sequence of bumps reflects the periodic fluctuations as function of $\log n$ analyzed in the upper bound. 

\begin{figure}
\includegraphics[width=8cm]{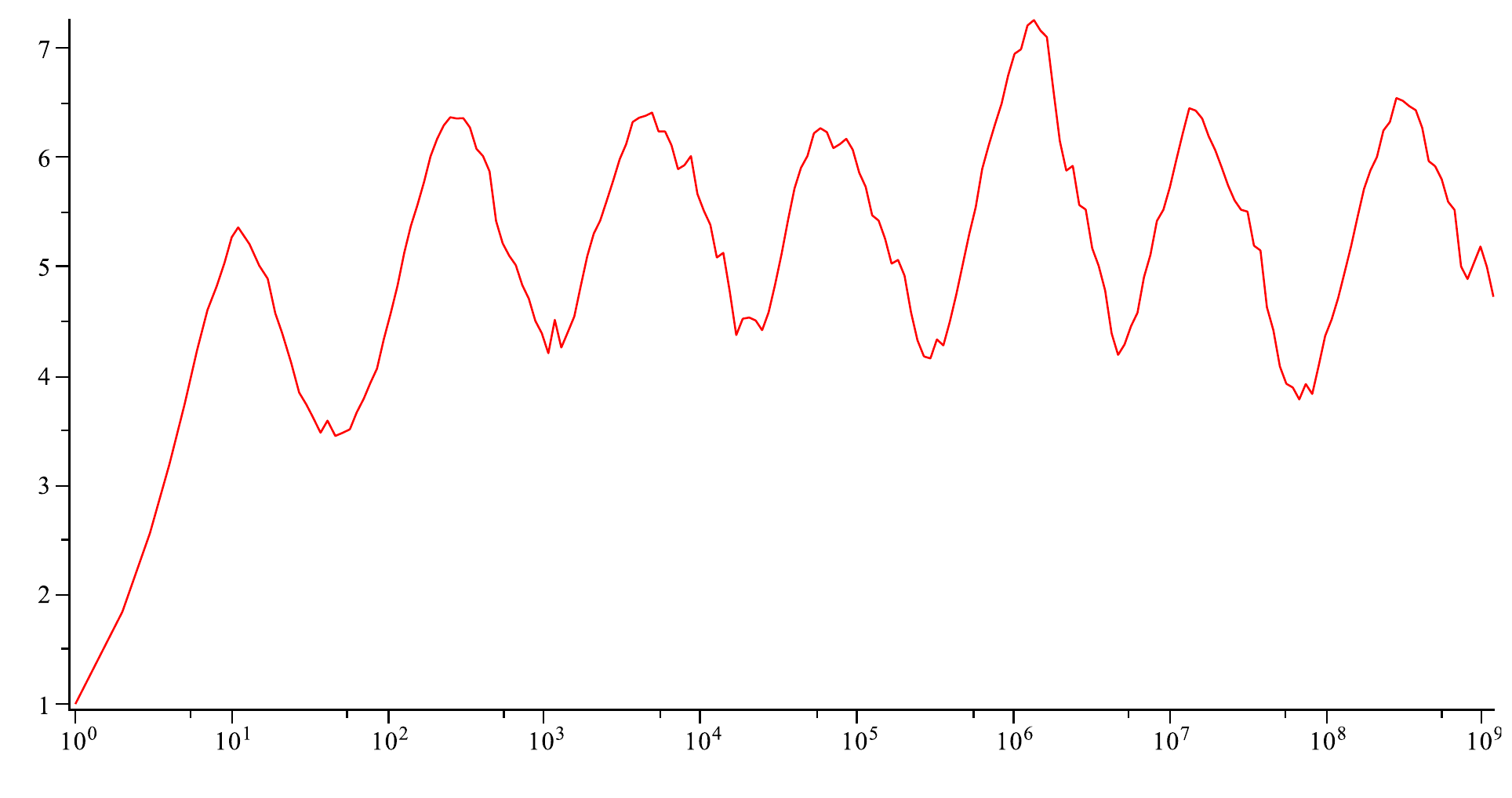}
\includegraphics[width=8cm]{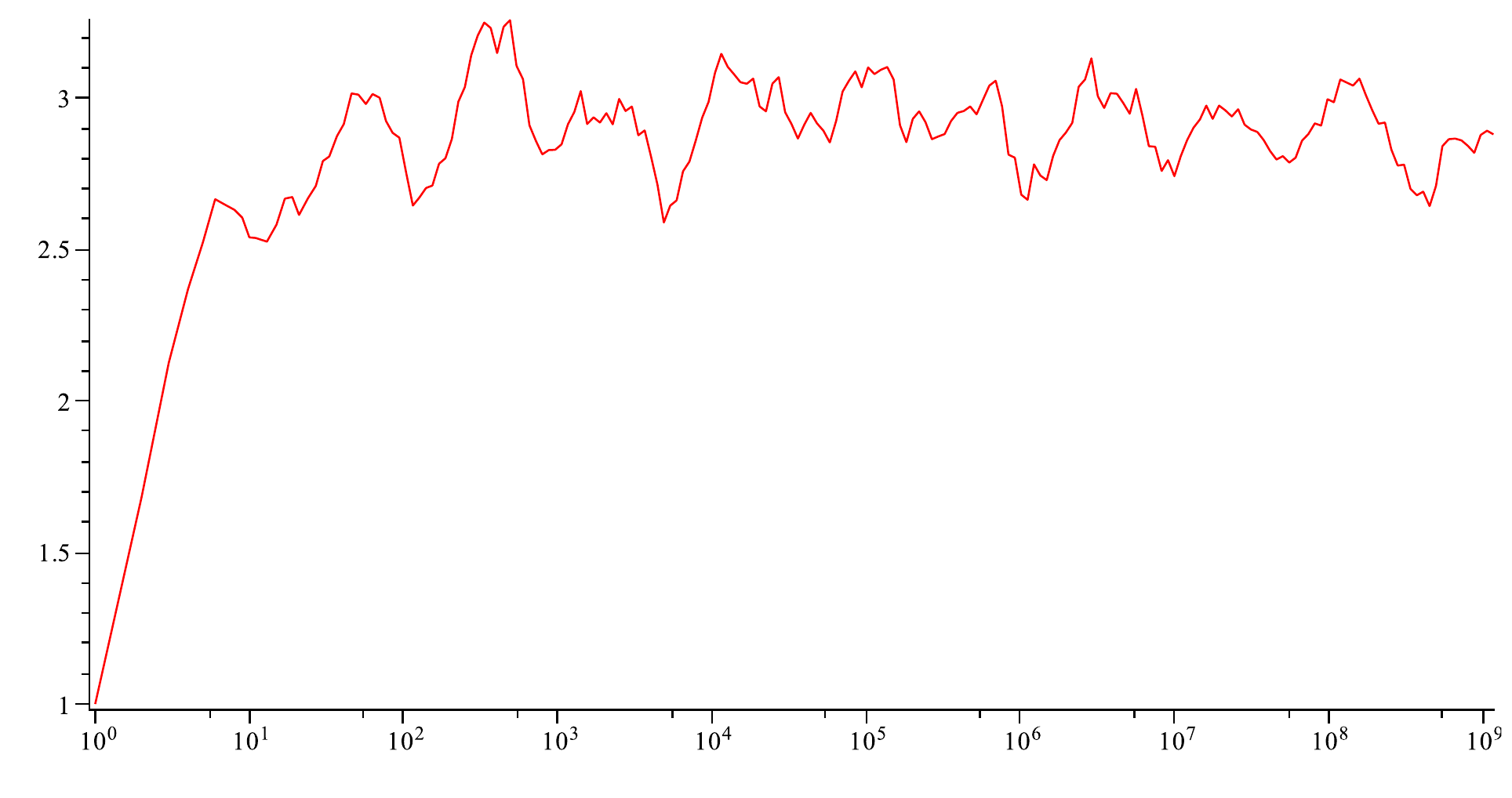}
\includegraphics[width=8cm]{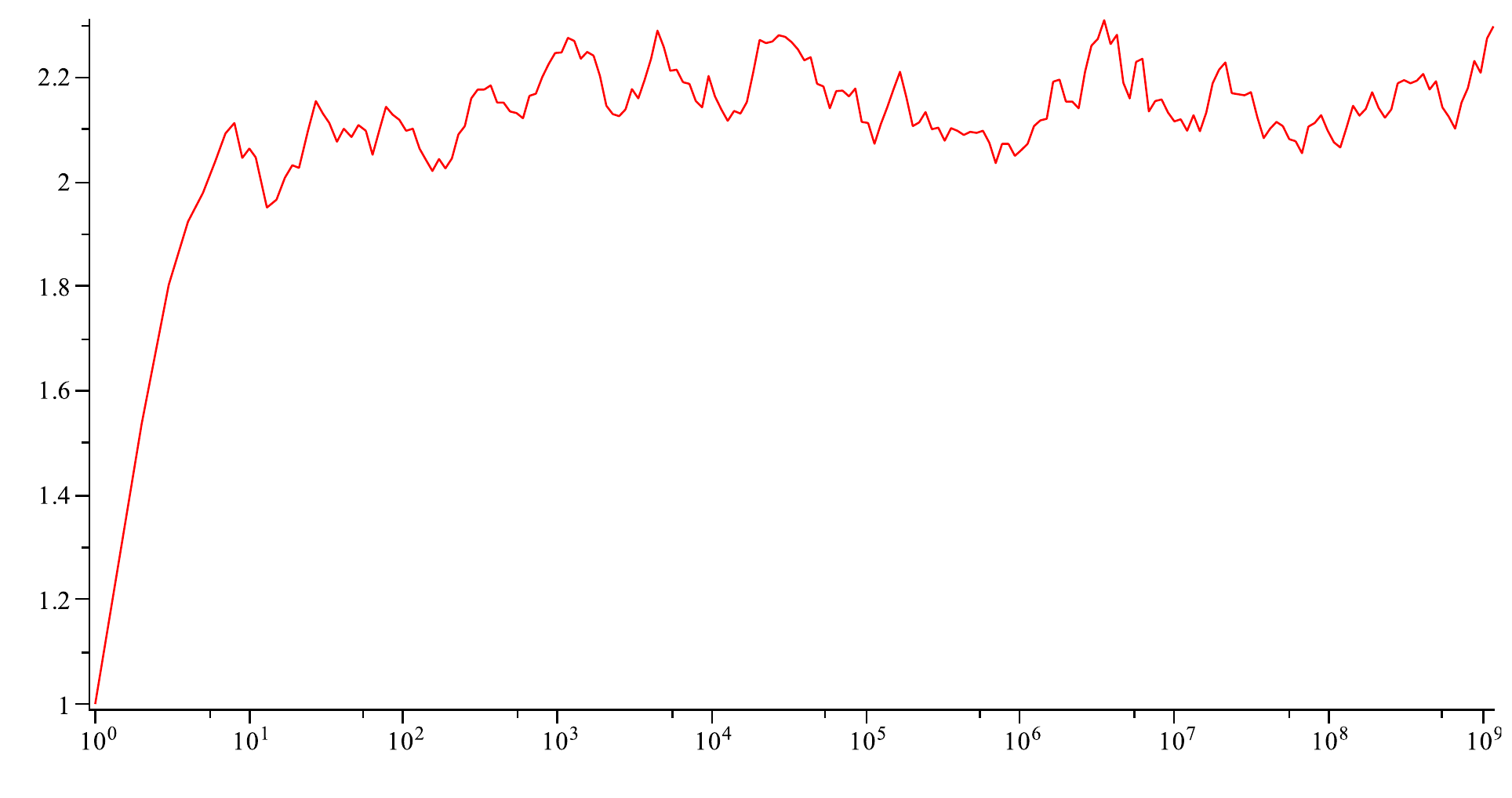}
\caption{The same as in figure~\ref{fig1} $E[\bM_n]$ with $N=2^{32}$ but obtained with 1,000 simulations per point, (top) for $k=8$, (middle) $k=12$, (bottom) $k=16$.}
\label{fig2}\end{figure}

Figure~\ref{fig2} shows the same quantities obtained by simulation, each point being simulated 1,000 times.

\begin{figure}
\includegraphics[width=8.5cm]{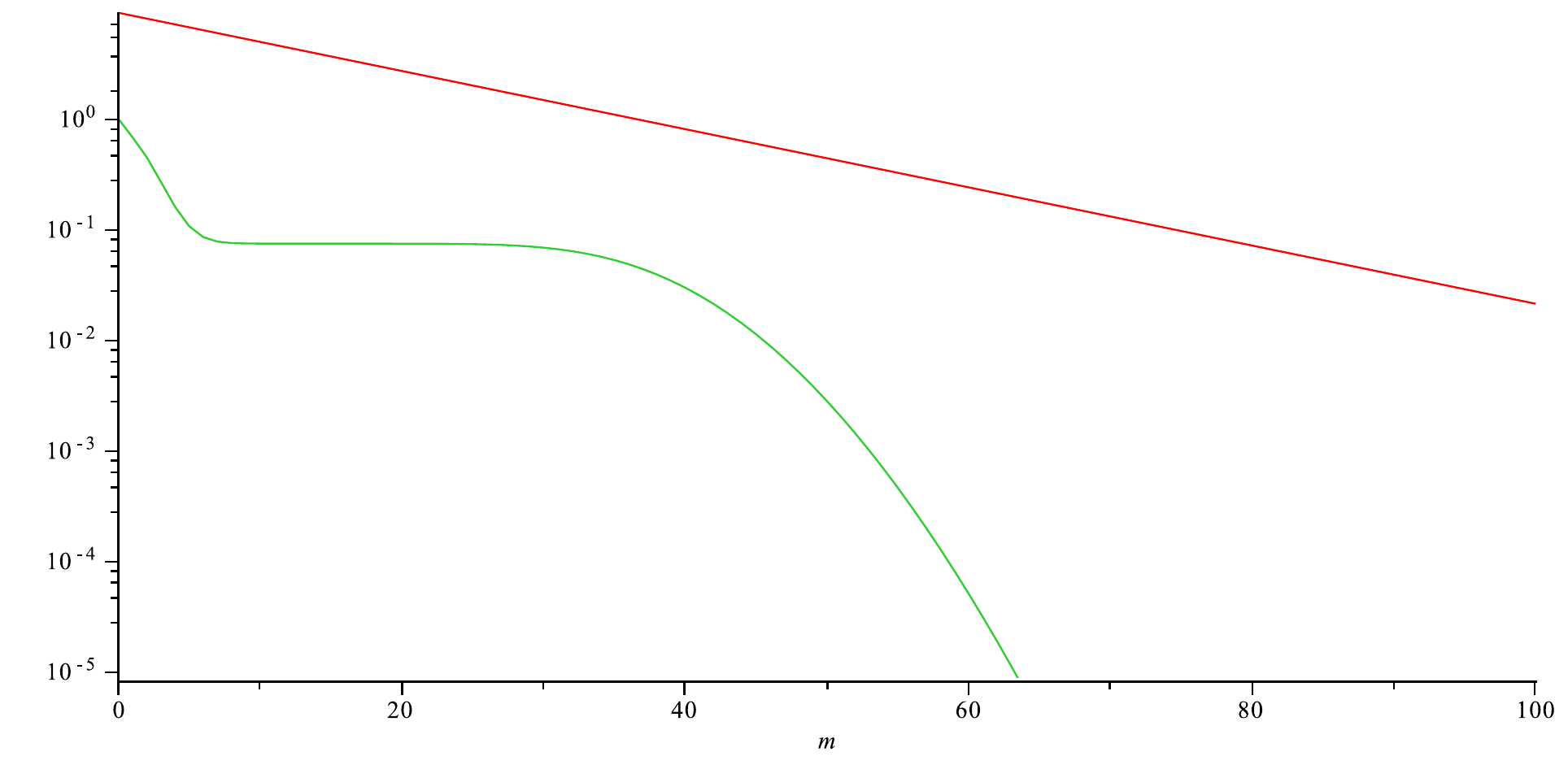}
\includegraphics[width=8.5cm]{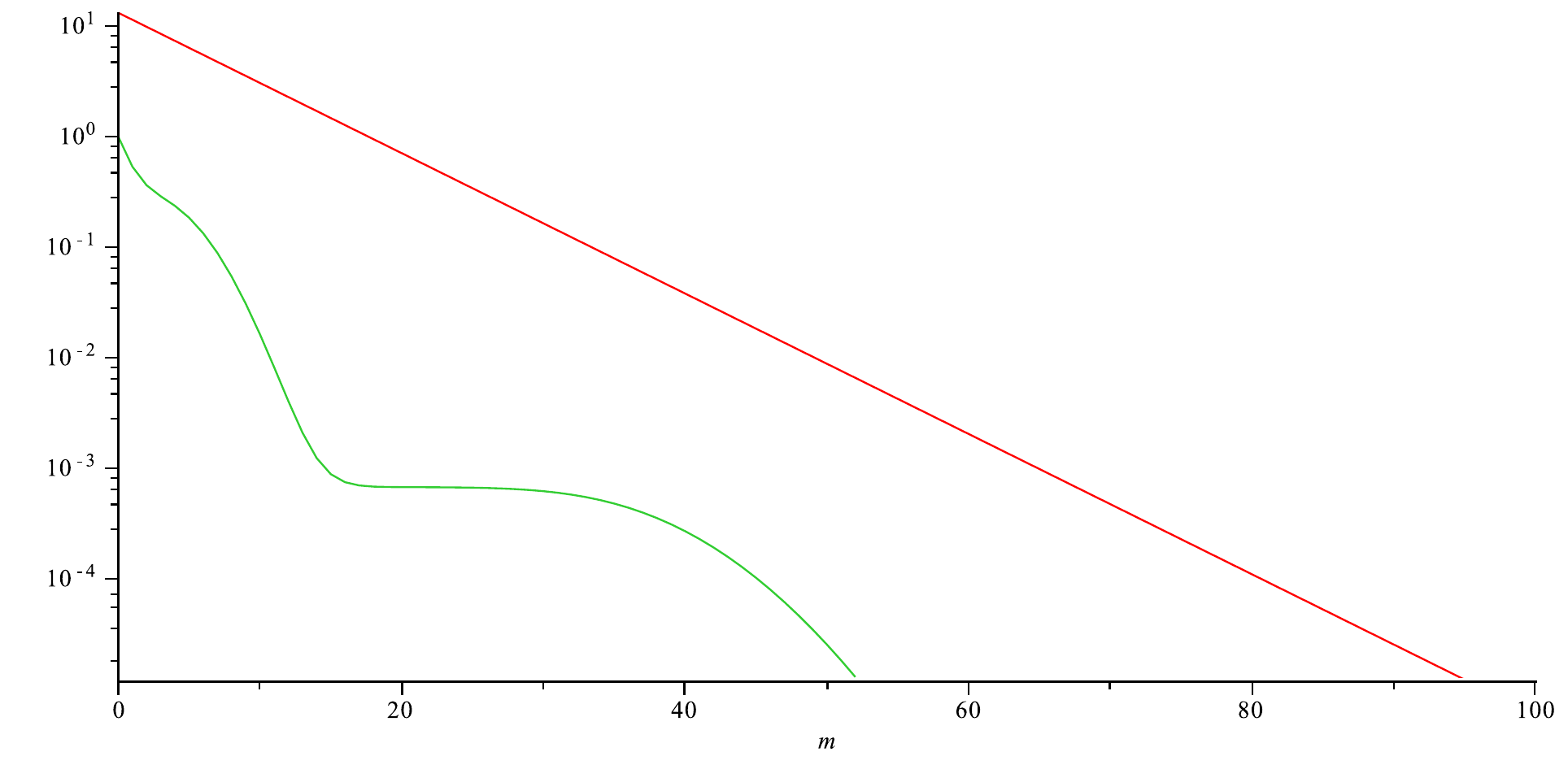}
\includegraphics[width=8.5cm]{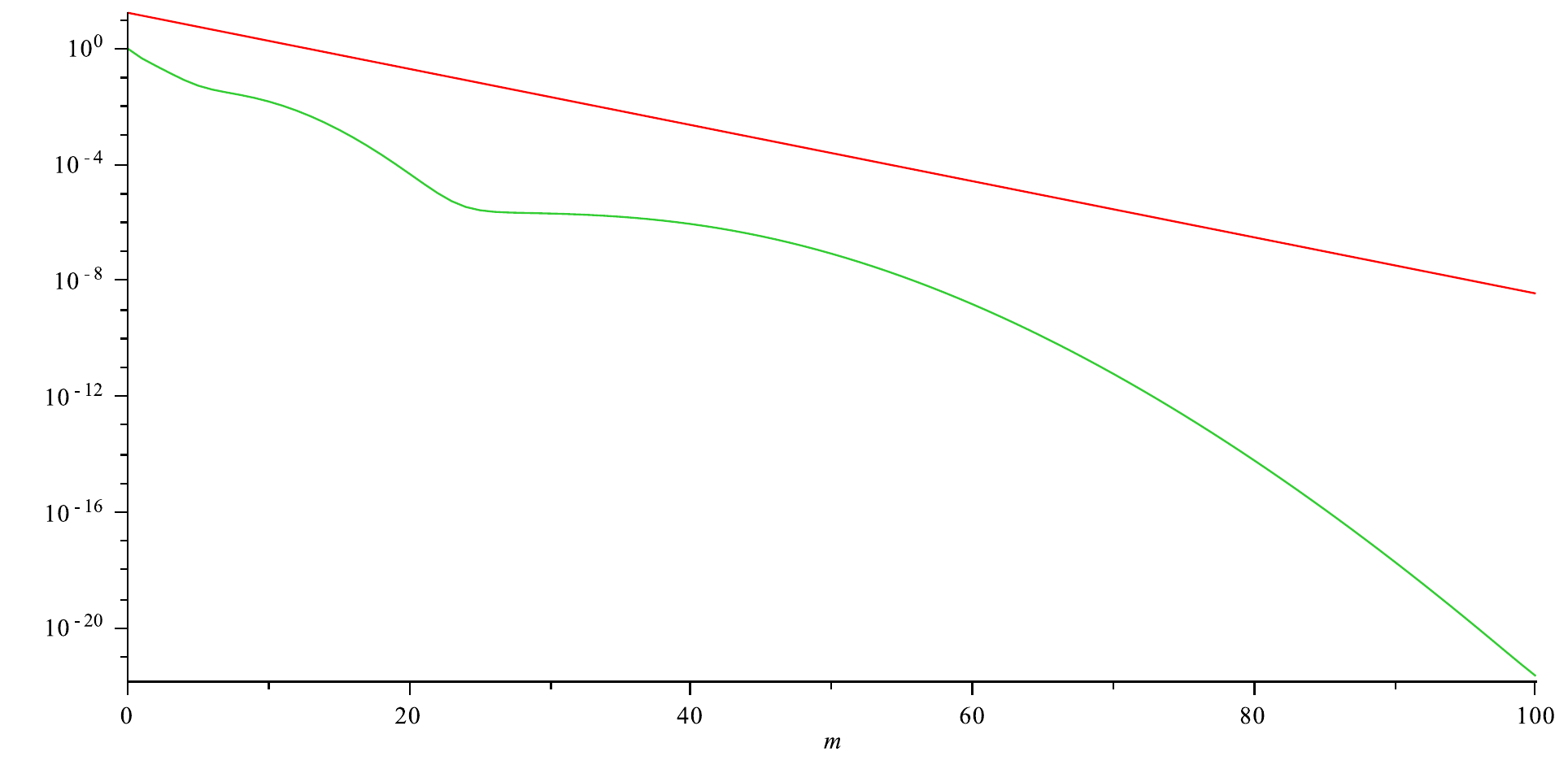}
\caption{Computed values of $P(\bM_n>m)$ with upperbound given by~(\ref{eq-Mnpu}) in red versus $m$ for $n=10,000$, with $N=2^{32}$, (left) for $k=8$, (right) $k=12$, (bottom) $k=16$.}
\label{fig4}\end{figure}

In fact, we can give a close expression of the distribution of $\bM_n$.
\begin{lemma}
For all complex number $u$:
\begin{eqnarray*}
E[u^{\bM_n}]&=&\left(1+P_0(u-1)\right)^n-(1-P_0)^n\\
&&+\sum_{0<\ell<k}\prod_{j<\ell}(1-P_j)^n\\
&&\times\left(\left(1+P_\ell(u-1)\right)^n-(1-P_\ell)^n\right)\\
&&+u^n\prod_{j<k}(1-P_j)^n.
\end{eqnarray*}
\end{lemma}
\begin{IEEEproof}
By application of Lemma~\ref{lemMnu}.
\end{IEEEproof}
Figure~\ref{fig4} displays the various shape of the sequences $P(\bM_n>m)$ versus integer $m$ for various parameters. Notice that the sequence of descending bumps in the plots comes from the mixture of the different binomial distributions.

\begin{figure}
\includegraphics[width=9cm]{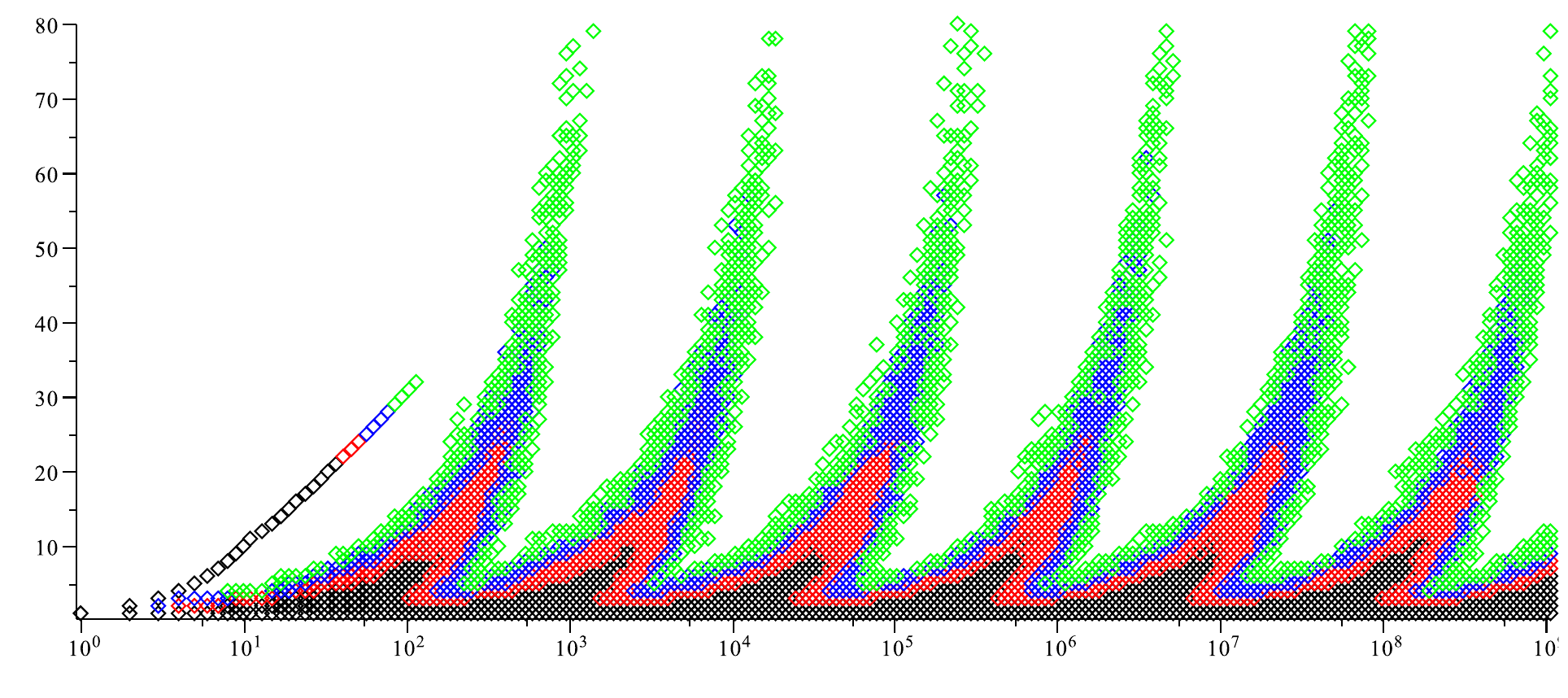}
\includegraphics[width=9cm]{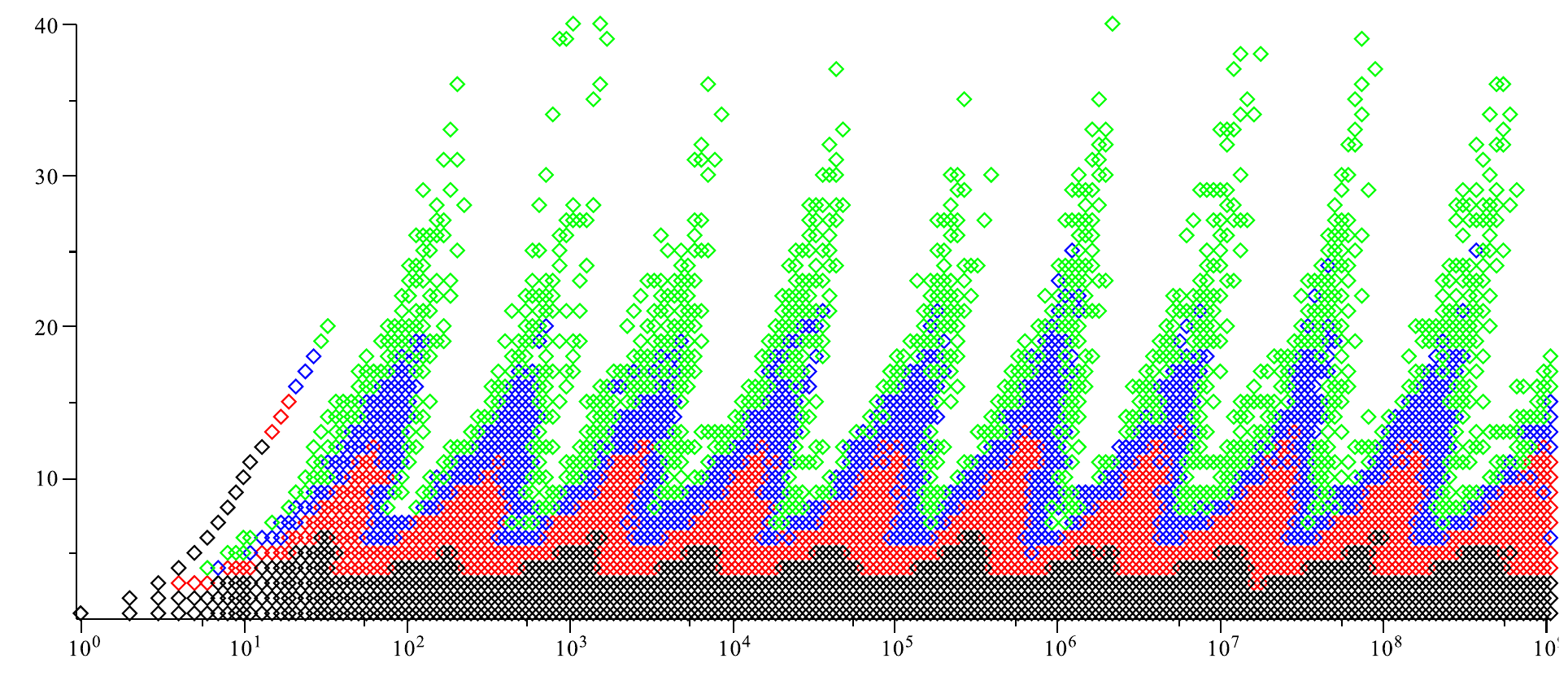}
\includegraphics[width=9cm]{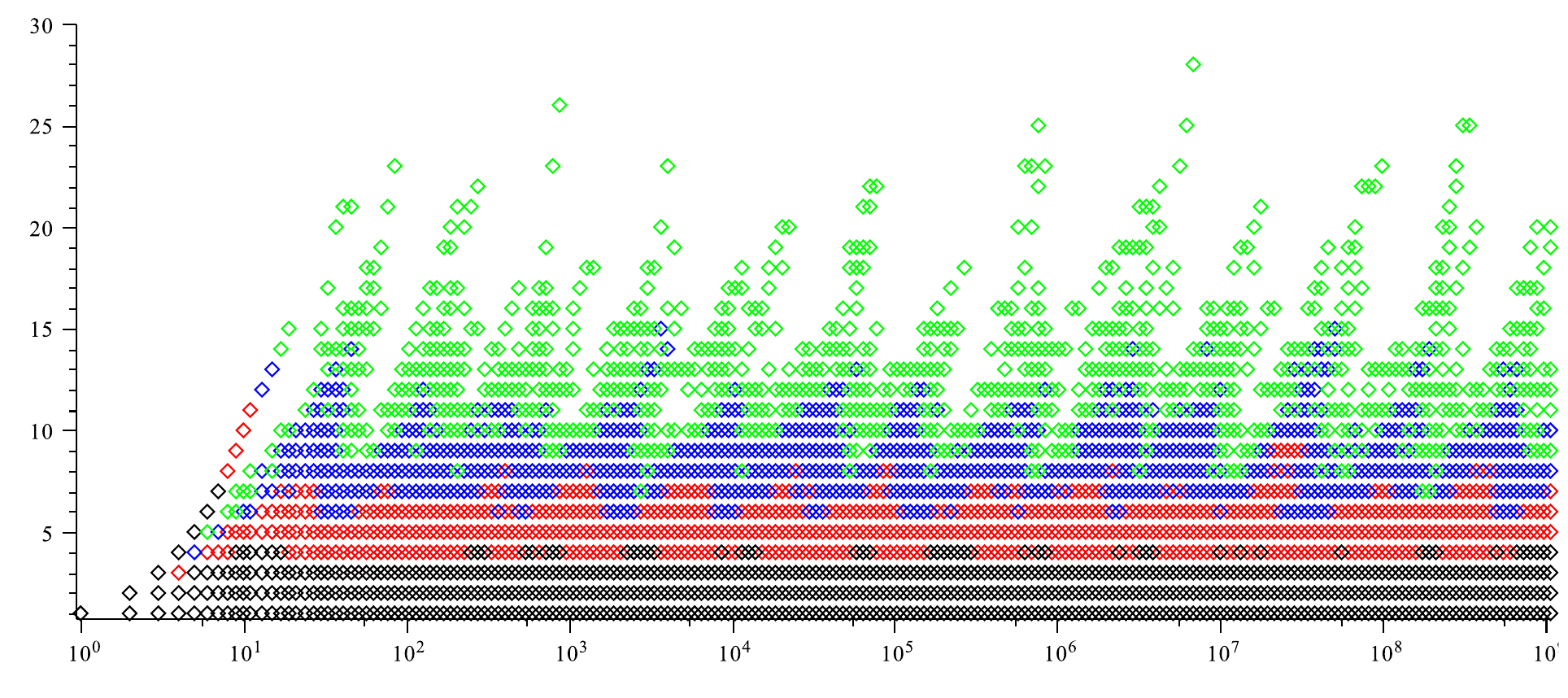}
\caption{various values of $\bM_n$ versus $n$ simulated 1000 times black values appeared more than 64 times, red more than 16 times, blue more than 4 times, green more than once, with $N=2^{32}$, (top) for $k=8$, (middle) $k=12$, (bottom) $k=16$.}
\label{fig3}\end{figure}

We can also give an estimate of the exponential tail distribution of $\bM_n$ by the following theorem about large deviations.
\begin{theorem}
Let $p=N^{-1/k}$. For all integer $m>0$, we have the estimate 
\begin{equation}
P(\bM_n>m)\le\frac{k+e^p}{(1+p)^m}
\label{eq-Mnpu}\end{equation}
\end{theorem}
\begin{IEEEproof}
We have $P_\ell=p^{k-\ell}$ and 
{\footnotesize
\begin{eqnarray*}
E[u^{\bM_n}]&\le&(1+p^k(u-1))^n-(1-p^k)^n+\\
&&\sum_{0<\ell<k}\left(\left(1+p^\ell(u-1)\right)^n-(1-p^\ell)^n\right)(1-p^{\ell+1})^n\\
&&+u^n(1-p)^n
\end{eqnarray*}
}
Let $u=1+p$, we have 
$$1+(u-1)p^\ell=1+p^{\ell+1}$$ and therefore 
$$(1+(u-1)p^\ell)(1-p^{\ell+1})=1-p^{2(\ell+1)}\le 1.$$ 
Thus $E[u^{\bM_n}]\le (1+p^{k+1})^n+k$. 

Since $(1+p^{k+1})^n\le\exp(np^{k+1})=\exp(pn/N)$ and $n<N$ we have 
$$\exp(np/N)\le\exp(N^{-1/k}).$$ 
We conclude the proof with the observation that for all integer $m$: $$u^mP(\bM_n>m)\le E[u^{\bM_n}].$$
\end{IEEEproof}
Figure~\ref{fig3} displays the histograms of 1,000 simulations of $\bM_n$ versus $n$ for $N=2^{32}$ and $k=8,12,16$. The color of the point $(n,m)$ indicates the number of times the event $\bM_n=m$ has been obtained during the simulation with the color code: black when more than 64 times, red when between 16 and 64, blue when between 4 and 16, green when between 1 and 4.


\section{Decentralized Empty block mining}
\label{sec:distributed}
For the sake of simplicity, we have initially described our scheme with a requirement to have a central authority to create and mine  empty blocks. We now remove this constraint by making the empty block creation fully distributed. For the sake of clarity, we extend our scheme in two steps (with the introduction of two features):
\begin{itemize}
\item (A) Time moderated empty blocks mining;
\item (B) implicit empty blocks mining.
\end{itemize}
\subsection{Time moderated empty block mining}
In this extended scheme, now all entities can mine empty blocks. In this case in order to avoid unfairness between empty miners, a new empty block will be allowed only if its date is above a minimal gap time between the last full and empty block time. For comparison with other existing schemes, the Minimal Gap Time (MGT) can be set to one minute (e.g., in Bitcoin this is called the Block Time, the average time between blocks, and the difficulty of the PoW is adjusted to make the time about 10 minutes; while the block time for Ethereum is set to between 14 and 15 seconds). Note that, in this case, we still need to have at least one entity mining empty blocks to guarantee liveness. Hence for this first variation we assume that possibly several (and at least one) peer-nodes  generates empty blocks and thus play the role of a distributed timestamping and block servers.

Time can be captured using Unix universal time (POSIX), a widely used time-stamping system in Unix-like and many other operating systems and file formats, that is commonly used for Blockchain systems. However, one difficulty is that the date given by the clock is not mandatorily accurate (e.g., Unix time is not a true representation of UTC, and leap seconds are not accounted for). However note we are only interested in the time difference in this part.
One possibility is that an additional time stamp is added to the block in order to reflect the local time when the timestamping server received the block. Of course, since the timestamping server's time is local, it can only be added at the reception and will be excluded from the hash value computation of the block. An empty block received before the expiration of the minimum time gap with the local time of the timestamping server  will be delayed. An empty block with a time stamp which does not show the MGT offset with the original time stamp of a full block will be discarded. 

Also the date field of the empty blocks must be excluded from their hash value computation. Otherwise full blocks with very different hash values could be called by different empty blocks on different peers just because the latter show different hash value due to different time stamps. Hence the sequence of hash values of the empty blocks mined after a full block will follow a deterministic sequence. If we ignore the clock drifts, the use of the time stamp does not change the performance analysis of the scheme as we retain the important property that two  consecutive calls to hash function generate independent hash values.

\subsection{Implicit empty block mining}
We can now extend the previous scheme by exploiting the fact that the sequence of hash values of the empty blocks mined after a full block will follow a deterministic sequence, and thus make them ``implicit''.
In addition, and finally, we remove completely the requirement of any special or centralized entity and make the scheme fully distributed: one does not need a special or central entity to initiate empty blocks since empty blocks will now become implicit, {\it i.e.} no empty blocks will be mined. The modification of the protocol is the following: 
\begin{itemize}
\item Upon reception of a full block within less than one MGT after the mining of the last full block, the block server discards it;
\item Upon reception of a full block within less than $\ell$ MGT ($1\le\ell\le k$) after the mining of the last full block, with either new block's hash value is larger than $ 2^h P_{\ell-1}$ or the previous block hash value is not the hash value of the last full block, the  block server discards it.
\end{itemize}

Note that another advantage of this scheme is that it also minimises the risk of creating forks as full blocks are discarded earlier.

\subsection{Performance analysis of the implicit empty block scheme}
\begin{figure}
\includegraphics[width=9cm]{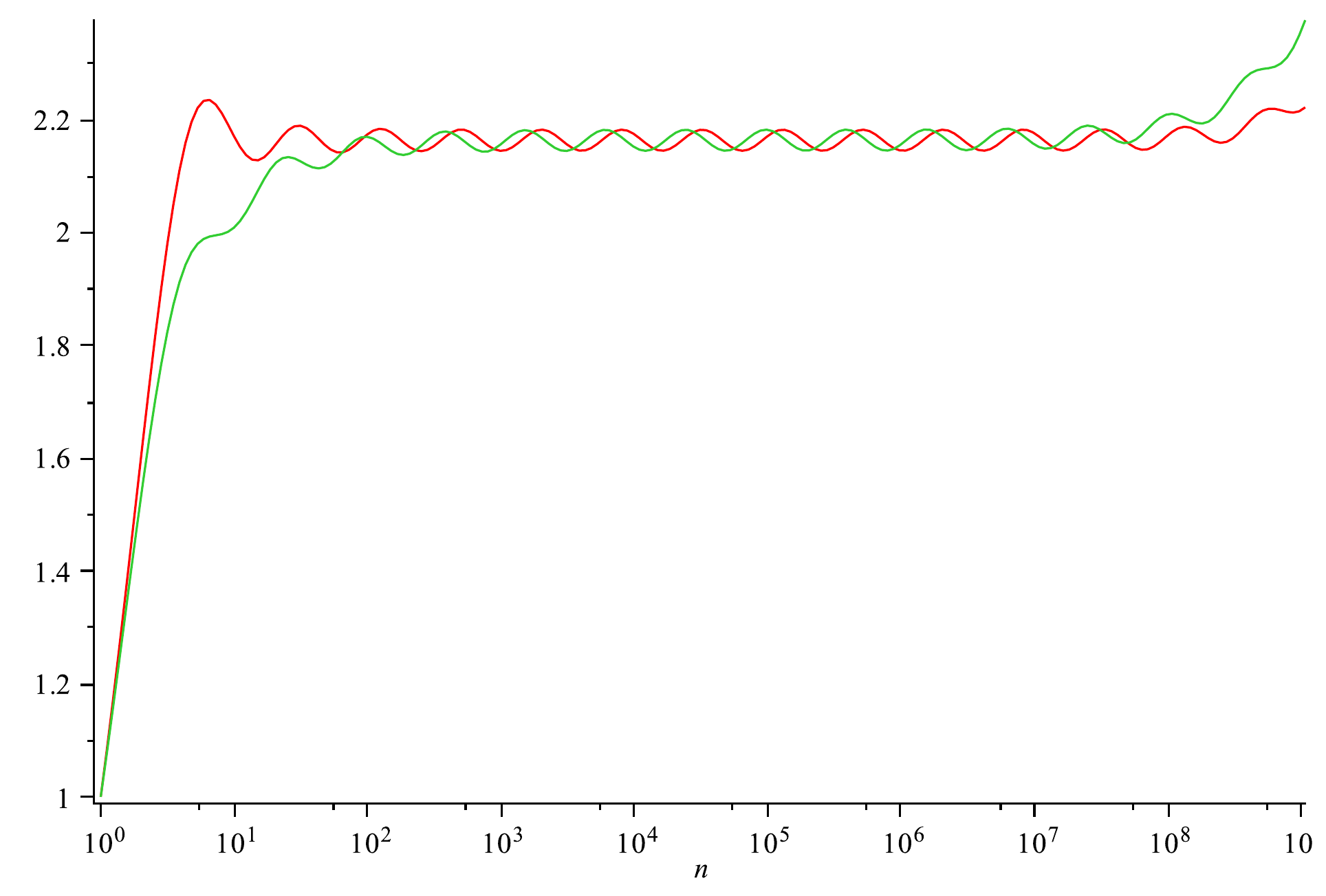}
\caption{Average number $E[\bM_n]$ of mined blocks with explicit (green) and implicit (red) empty blocks mining versus $n$ for $k=16$.}
\label{figimplicit}\end{figure}

The main difference with the previous scheme analysis is that the hash value of the blocks after each empty block are no longer independent. In fact the hash value of a full block candidate is no longer possibly related to a new empty block. Indeed the previous block hash value field is no longer the hash value of the previous empty block, since it does no longer exists, but the hash value of the last full block. 
Keeping the previous notations we can state our main theorems and lemmas.
\begin{lemma}
For $m>0$ we have the expression
\begin{eqnarray*}
P(\bM_n=m)&=&\binom{n}{m}P_0^m(1-P_0)^{n-m}\\
&&+\sum_{\ell=1}^{k-1}\binom{n}{m}(P_{\ell}-P_{\ell-1})^m(1-P_\ell)^{n-m}\\
&&+\delta(m-n)(1-P_{k-1})^n
\end{eqnarray*}
\end{lemma}
\begin{IEEEproof}
We call \emph{slot} the time interval of length MGT. Let assume that $n$ full blocks are in competition. 
Let $0<\ell<k$, the probability that a block has a hash value comprised between $2^hP_{\ell-1}$ and $2^hP_{\ell}$ is $P_{\ell}-P_{\ell-1}$. Such a block will be mined on the $\ell+1$-th slots after the last full block if and only if all the other full blocks in competition have their hash values greater than or equal to $2^hP_\ell$. Those which have the hash value greater than $2^hP_\ell$ will not be mined. The hash values of the blocks being assumed independent justifies the binomial expression. The case $\ell=0$ corresponds to blocks which have value smaller than or equal to $2^hP_0$. In the case $\ell=k$ we have $P_k=1$ and occur when all the blocks have hash values greater than or equal to $2^hP_{k-1}$
\end{IEEEproof}
\begin{lemma}
We have the expression:
\begin{equation}
E[\bM_n]=nP_0+\sum_{\ell=1}^{k}n(P_\ell-P_{\ell-1})(1-P_{\ell-1})^{n-1}.
\end{equation}
\end{lemma}
\begin{IEEEproof}
This is a direct expression of the previous lemma. 
We have $E[\bM_n]=\frac{\partial}{\partial u}E[u^{\bM_n}]|_{u=1}$. Indeed for $u$ complex:
\begin{eqnarray*}
\sum_{m}u^m\binom{n}{m}(P_\ell-P_{\ell-1})^m(1-P_{\ell-1})^{n-m}=\\
((P_\ell-P_{\ell-1})u+1-P_{\ell-1})^n
\end{eqnarray*}
and the derivative of the right hand side with respect to variable $u$ is equal to $n(P_\ell-P_{\ell-1})((P_\ell-P_{\ell-1})u+1-P_{\ell})^{n-1}$.
We notice that when $\ell=k$ we have 
$$n(P_\ell-P_{\ell-1})(1-P_{\ell-1})^{n-1}=n(1-P_{k-1})^n$$ 
since $P_k=1$.
\end{IEEEproof}
The performance of the implicit empty blocks strategy are comparable to the performance of the explicit empty blocks. In fact we see that the plots of the average number of mined blocks oscillate and cross each other when we compare both strategy (see Figure~\ref{figimplicit}).

\begin{theorem}
For $n\le N$ we have the property: $$E[\bM_n]=O(N^{1/k}).$$
\end{theorem}
\begin{IEEEproof}
We have 
$$E[\bM_n]-P_0 n=\sum_{\ell=1}^kn(P_\ell-P_{\ell-1})(1-P_\ell)^{n-1}.$$ 
Since $P_\ell=p^{k-\ell}$ and $p^k=1/N$ we get
\begin{eqnarray*}
E[\bM_n]-\frac{n}{N}&=&\sum_{\ell=1}^{k}n (1/p-1)P_{\ell}(1-P_{\ell-1})^{n-1}\\
&=&\sum_{\ell=1}^{k}n (1/p-1)P_{\ell}\frac{(1-P_{\ell-1})^{n}}{(1-P_{\ell-1})}\\
&\le&\frac{1}{p}\sum_{\ell=1}^{k}nP_{\ell}(1-P_{\ell-1})^{n}
\end{eqnarray*}
since $(1-P_{\ell-1})^{-1}\le(1-p)^{-1}$. 

We recognize in the last right hand term an expression which we have already proven to be $O(N^{1/k})$ in the proof of Theorem~\ref{theo1}.
\end{IEEEproof}
Using the same reasoning, we obtain a more precise estimate via the use of the "Gamma" function: 
$$E[\bM_n]\le\frac{n}{N}+BN^{1/k}$$ 
with 
\begin{equation}
B=\frac{1}{p\log(1/p)}\sum_{k\in\mathbb{Z}}\left|\Gamma\left(1+\frac{2ik\pi}{\log p}\right)\right|.
\end{equation}

\section{Game theoretic approach for block nursing, 51 percent attack}
\label{blocknursing}
\subsection{The "51 percent" attack}
With the nonce disappearing in our scheme (from the internal format of the block that is hashed), any Proof of Work farming is no longer possible in order to achieve a competitive advantage in the mining contest. Indeed to change the hash function of the block, there is a need to change the content of the block, {\it i.e.} the Merkel tree representing the transaction confirmed by the block. In order to make the task even more difficult one can enforce the rule that the Merkel tree represents an ordered list of transactions. Under this condition modifying the hash function would need to modify the set of transactions contained in the block. This would make block nursing far less easy than proof of work farming since transaction identifiers should be distributed in caches while proof of work only require local increments of the nounce integer.

Nevertheless the proof of work farming could be replaced by {\em block nursing}. Instead of working on a single block and changing its hash function via the nonce, an institution prepares a set of blocks with different transaction combinations, since it is the only way to modify the block hash function. 

Nursing blocks will have a higher cost per hash function than with Proof of Work farming, since one has to access and retrieve transaction IDs. Furthermore not every transaction combination may be permitted, and some may not be interesting, resulting into an unacceptably low income. Also the set of currently available transactions may not be large enough to offer enough combinations. However, in the following we will assume that we have no such limit and assess the risk of nursing blocks. 

In this section we analyse the so-called 51 percent attack. It consists to measure the additional means in terms of processing power block nurse must commit in order to gain one term $\epsilon$ in the probability of prevailing against an adversary nurse.

We will prove the following theorem:
\begin{theorem}
if the adversaries are of equivalent power then one adversary increases of $\epsilon$ of its probability to win a block mining needs an increase of $\frac{1-p^2}{2p\log(1/p)}\epsilon$ of the size of block nurse factory.
\label{theo51}\end{theorem}
We notice that if $p$ is small the amplifying factor can be large. for $p=2^{-32/k}$ for $k=8$, $12$, $16$ we respectively get an amplification factor of $6.64$, $3.95$, $3.32$.

\subsection{Detailed analysis}
Let $A$ be a competitor nursing $m$ blocks and $B$ another competitor $B$ nursing $n$ blocks, let $L_{m,n}$ be the probability $L_{m,n}$ of $A$ losing for a block mining against $B$. We will show that the quantity $L_{m,n}$ starts to decrease significantly only when $m$ is of the order $n/p$. Since in the above examples one has $1/p$ greater than $4$ or more, this would require a major investment compared to $B$. In reciprocity if $B$ wants to balance with $A$, it would be sufficient that $B$ invests a little more than $p$ times $A$'s investment. Furthermore $1/L_{m,n}$ is the average length of consecutive mined blocks of miner $A$ against miner $B$. It estimates the ability of miner $A$ to have committed blocks and fork from miner $B$ to take the lead on the currency.

We give the following theorem in the context of {\it explicit} empty block mining, where the consecutive hash values are assumed to be independent random variables.

\begin{theorem}
We have $L_{m,n}$ asymptotically larger than $L(x)$ where $x=m/n$ and
$$
L(x)=\frac{\log((x+p)p)-\log(px+1)}{2\log p}(1+Q(\log x))
$$
with $Q(.)$ a periodic function of small amplitude and period $\log(1/p)$. 
\end{theorem}
\begin{IEEEproof}
We provide the main steps of the proof. After each non empty mined block there is a potential sequence of $k$ slots, called with empty blocks, to insert the next non empty block. If the calls make that $A$ inserts a block on a slot before the slot of $B$, $A$ wins the mining. If $A$ and $B$ select the same slot, $A$ and $B$ either win or loose with probability $1/2$. (Note that for the sake of the explanation here, and without loss of generality, we assume  that $A$ and $B$ mine only one block each.)

If we assume that the all hash function for the blocks nursed by $A$ are independent, the probability that  $A$ mines a block on the first slot is $1-(1-P_0)^m$, the probability that it mines a block on slot $\ell$ is $(1-P_0)^m\times\cdots\times(1-P_{\ell-1})^m(1-(1-P_\ell)^m)$. At the same time the probability that the competitor $B$ mines a block on slot $\ell$ is also $(1-P_0)^{n}\cdots(1-P_{\ell-1})^{n}(1-(1-P_\ell)^n)$. The probability that a competitor mines a block on a further slot is $(1-P_0)^{n}\cdots(1-P_{\ell-1})^{n}(1-P_\ell)^n$. Therefore the probability that player $A$ mines a block at slot $\ell$ and wins against $B$ is equal to the sum of half of the first probability plus the second probability, namely $\frac{1}{2}(1-P_0)^{n+m}\cdots(1-P_{\ell-1})^{n+m}(1-(1-P_\ell)^n)(1+(1-P_\ell)^m)$:
\begin{eqnarray*}
L_{m,n}&=&\frac{1}{2}(1-(1-P_0)^n)(1+(1-P_0)^m)\\
&&+\frac{1}{2}\sum_{\ell=1}^k(1-P_0)^{n+m}\cdots(1-P_{\ell-1})^{n+m}\\
&&\times(1-(1-P_\ell)^n)(1+(1-P_\ell)^m)\
\end{eqnarray*}

Thus we get
\begin{eqnarray*}
L_{m,n}&=&\frac{1}{2}\sum_{\ell=0}^k(1-(1-p^\ell)^n)(1+(1-p^\ell)^m)\\
&&\times\prod_{i>\ell}^k(1-p^i)^{n+m}\\
&\ge&\frac{1}{2}\sum_{\ell=0}^k(1-(1-p^\ell)^n)(1+(1-p^\ell)^m)\\
&&\times(1-\frac{p^{\ell+1}}{1-p})^{n+m}
\end{eqnarray*}
Since $p^k=1/N$, we have
$$
L_{n,m}\ge L_x(n)+O(n/N).
$$
with $$L_x(n)=\frac{1}{2}\sum_{\ell\ge0}(1-(1-p^\ell)^n)(1+(1-p^\ell)^{nx})(1-\frac{p^{\ell+1}}{1-p})^{(1+x)n}.$$
The Mellin transform of $L_x(n)$ with respect to variable $n$ is $$L_x^*(s)=\frac{\Gamma(s)}{2}\sum_{i=0}^1\sum_{j=0}^1(-1)^iL_x^{i,j}(s)$$ with 
{\small
$$
L^{i,j}_x(s)=\sum_{\ell\ge 0}\left(-(1+x)\log(1-\frac{p^{\ell+1}}{1-p})-(i+jx)\log(1-p^\ell)\right)^{-s}
$$
}
since $L^{i,j}_x(s)=\frac{\left((1+x)\frac{p}{1-p}+(i+jx)\right)^{-s}}{1-p^{-s}}+O(\Re(s))$. 

The function $L^*(s)$ has therefore a pole at a $s=0$ and at integer multiples of $2i\pi/\log p$. The pole at $s=0$ gives a contribution to the asymptotic of $L_x(n)$, when $n\to\infty$, which is 
$$
\frac{\sum_{i=0}^1\sum_{j=0}^1(-1)^i\log\left((1+x)\frac{p}{1-p}+(i+jx)\right)}{2\log p}=L(x)
$$
The other poles contributes to the periodic term $Q(\log x)$.
\end{IEEEproof}
We notice that 
\begin{eqnarray*}
L(1)&=&1/2,\\ 
L(1/p)&=&\frac{\log(1+p^2)-\log 2}{2\log p}\\
&\approx&\frac{\log 2}{2\log(1/p)},\\
\lim_{x\to 0}L(x)&=&1, \\
 \lim_{x\to\infty}L(x)&=&0, \\
L(x) &\sim& \frac{p-1/p}{2x\log p}, \textnormal{when } x\to\infty,\\
L(x)&=&1+\frac{x}{2p\log p}+O(x^2),
\textnormal{when } x\to 0.
\end{eqnarray*}
The last identity proves the 51 percent result of Theorem~\ref{theo51}
\begin{figure}
\includegraphics[width=9cm]{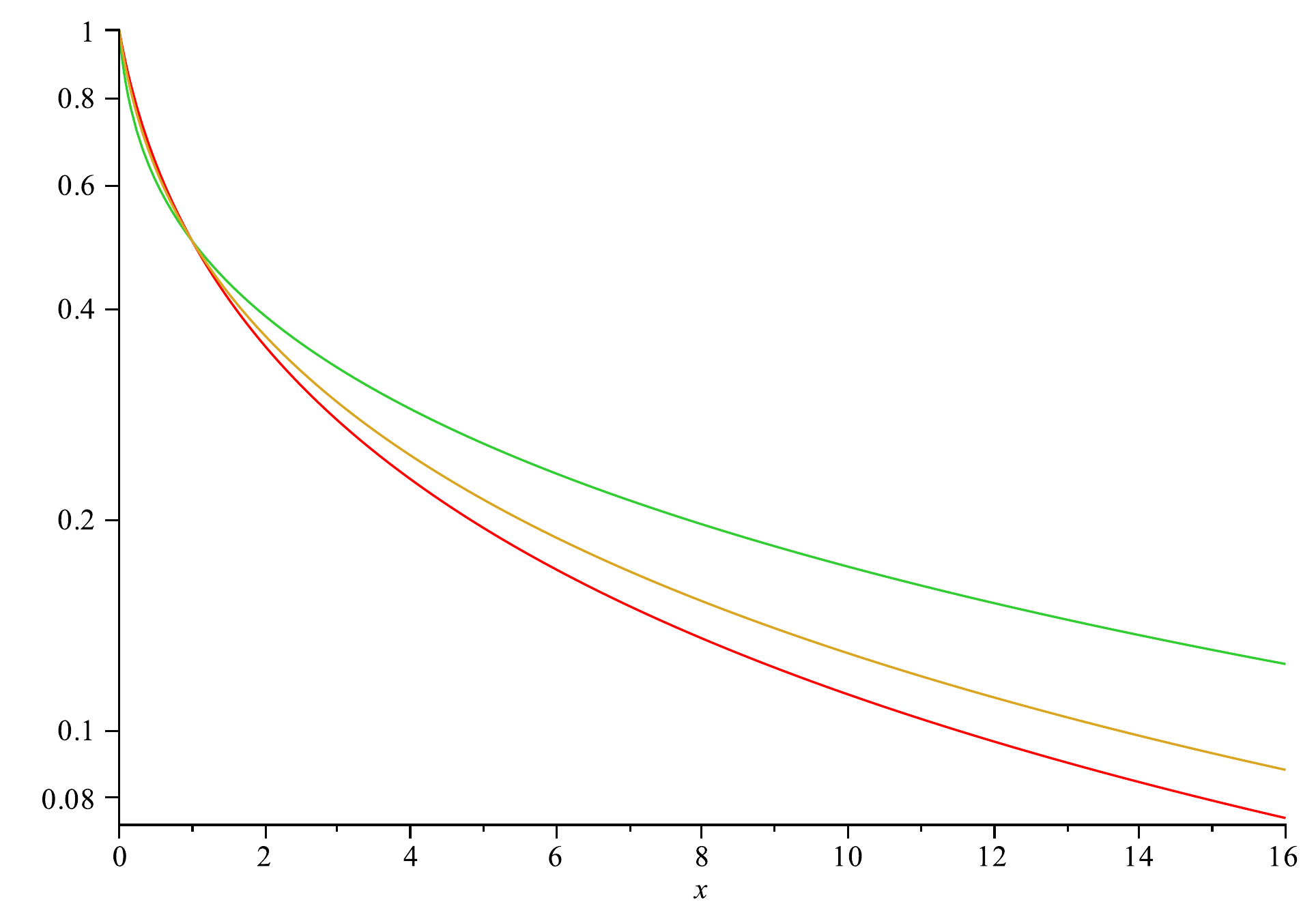}
\includegraphics[width=9cm]{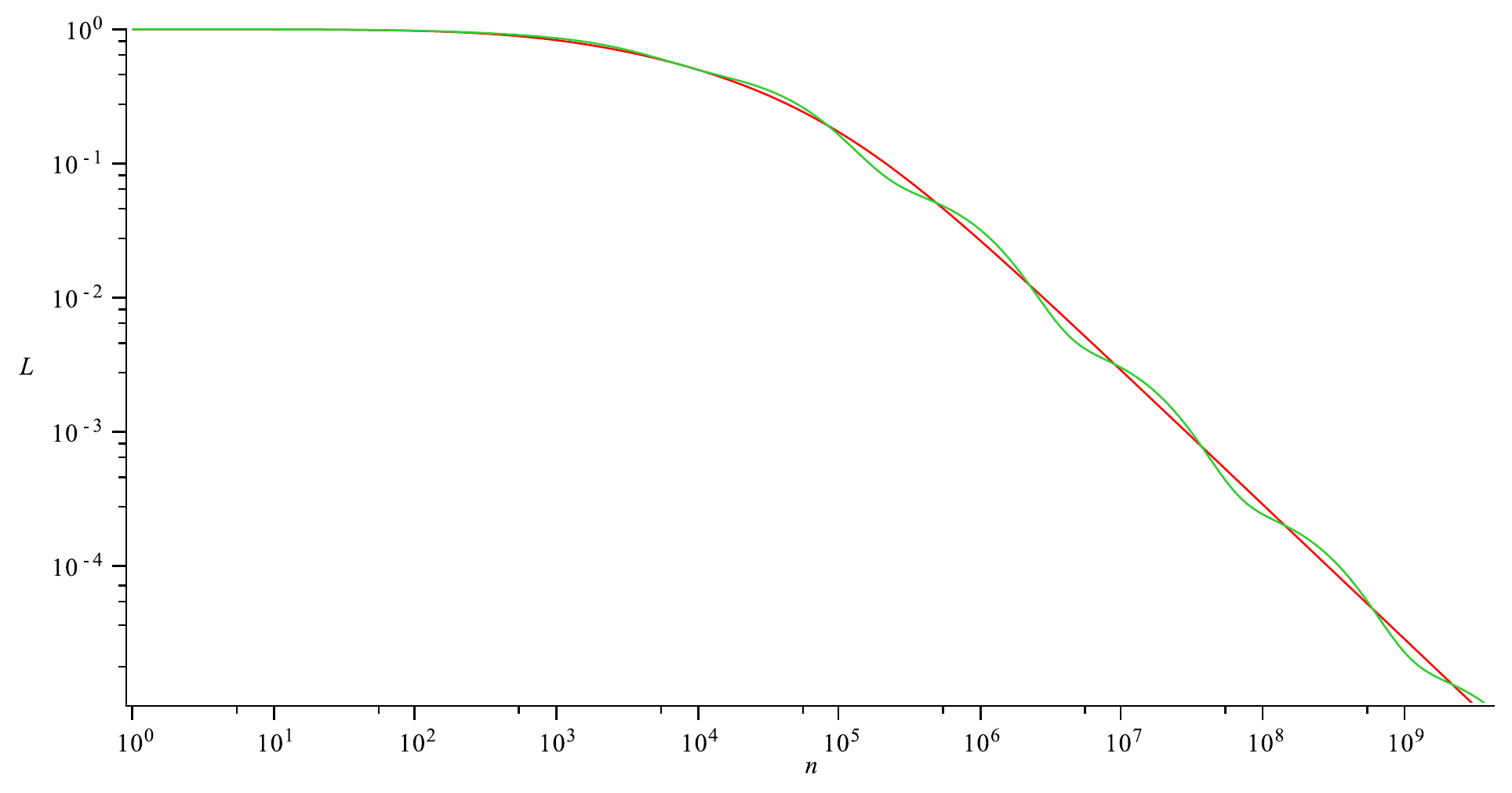}
\caption{Function $L(x)$ (up), function $L_{m,n}$ for $n=10,000$ (down), both with $N=2^{32}$ and $k=8$.}
\label{figloss}\end{figure}

\section{Conclusion}
\label{conclusion}
We have presented a new blockchain scheme which gets rid of the overwhelming waste of energy created by the Proof of Work used in Blockchain protocols such as Bitcoin. The technique used for the proof of our scheme is based on the analysis of a distributed Green Leader Election (which is of interest on its own). 

The main difference with Bitcoin-like Proof-of-Work schemes beyond the suppression of the nonce field (triggering most of the waste), is the introduction of empty blocks which aims are to call regular blocks following a controlled staircase set of values. In a simplified approach (that can be relevant for some particular sets of applications or peers), the security of the scheme comes from a secured repository of mined empty blocks by a single authority. Hence we also introduced and analysed a completely distributed block mining protocol that uses an implicit empty block mining which does not need any central authority. 

With the disappearance of the nonce in the proof of work, we also consider attacks on the cryptocurrency where it would take other variants of proof of work farming. We analysed the performance of a new strategy, which we name "block nursing" to prove that it will be inherently more complicated than proof of work farming and much less rewarding. 

\section*{Acknowledgment}
For the research, B.M. was partially supported by the Australian Research Council Grants DP140100118 and DP170102794. Part of this work was done when BM was visiting LINCS.

We are indebted to Daniel Augot for earlier discussions on Blockchain systems.



\begin{thebibliography}{9}

  
  





\bibitem{aspnes03}
James Aspnes,
\emph{Randomized protocols for asynchronous consensus}, Distributed Computing, 16 (2-3):165–175, 2003.

\bibitem{aspnes06}
James Aspnes, Collin Jackson, and Arvind Krishnamurthy, \emph{Exposing computationally challenged Byzantine impostors}, Technical Report YALEU/DCS/TR-1332, Yale University Department of Computer Science, July 2005.

\bibitem{Useful2017}
Marshall Ball, Alon Rosen, Manuel Sabin, and Prashant Nalini Vasudevan. \emph{Proofs of useful work}, IACR Cryptology ePrint Archive 2017: 203 (2017).

\bibitem{Ball2017}
Marshall Ball, Alon Rosen, Manuel Sabin, and Prashant Nalini Vasudevan. \emph{Average-case fine-grained hardness}, in proceedings of the 49th Annual ACM SIGACT Symposium on Theory of Computing, (STOC), Montreal, QC, Canada, June 19-23, 2017, pages 483–496. ACM, 2017.

\bibitem{Ball2018}
Marshall Ball, Alon Rosen, Manuel Sabin, Prashant Nalini Vasudevan, 
\emph{Proofs of Work from Worst-Case Assumptions}, IACR Cryptology ePrint Archive 2018: 559 (2018).

\bibitem{Activity2014}
Iddo Bentov, Charles Lee, Alex Mizrahi, and Meni Rosenfeld, \emph{Proof of activity: Extending bitcoin’s proof of work via proof of stake}. ACM SIGMETRICS Performance Evaluation Review, 42(3):34–37, 2014.
  
\bibitem{Consensus2017}
Tyler Crain, Vincent Gramoli, Mikel Larrea, Michel Raynal,
\emph{Leader/Randomization/Signature)-free Byzantine Consensus for Consortium Blockchains}, CoRR abs/1702.03068 (2017)

\bibitem{decker13}
Christian Decker and Roger Wattenhofer,
\emph{Information Propagation in the Bitcoin Network},
in proceedings of the 13th IEEE International Conference on Peer-to-Peer Computing (P2P), Trento, Italy, September 2013.

\bibitem{dwork92}
Cynthia Dwork and Moni Naor,
\emph{Pricing via processing or combatting junkmail},
in proceedings of the 12th Annual International Cryptology Conference on Advances in Cryptology (CRYPTO), pp.~139–147, London, UK, 1993.

\bibitem{garay15}
Juan Garay and Leonardos Kiayias, 
\emph{The bitcoin backbone protocol: Analysis and applications}, in proceedings of the 34th Annual International Conference on the Theory and Applications of Cryptographic Techniques (EUROCRYPT), pp.~281–310, 2015.

\bibitem{garay17}
Juan Garay, Leonardos Kiayias and N. Leonardos 
\emph{The bitcoin backbone protocol with chains of variable difficulty} in proceedings of the 36th Annual International Cryptology Conference on Advances in Cryptology (CRYPTO), LNCS, vol. 10401, pp. 291–323. Springer, 2017.

\bibitem{Grabner}
P.J.~Grabner and H.~Prodinger, 
\emph{Sorting algorithms for broadcast communications: Mathematical analysis}, Theoretical computer science, 289(1), pp.~51-67, 2002.

\bibitem{green}
Philippe Jacquet, Dimitris Milioris and Paul M{\"{u}}hlethaler,
\emph{A Novel Energy Efficient Broadcast Leader Election},
In proceedings of the 21st International Symposium on Modelling, Analysis and Simulation of Computer and Telecommunication Systems (MASCOT), San Francisco, August, 2013, pp.~495--504. 

\bibitem{cichon} Cichon, J., Kapelko, R., \& Markiewicz, D. (2016). \emph{On Leader Green Election}. In proceedings of the 27th International Conference on Probabilistic, Combinatorial and Asymptotic Methods for the Analysis of Algorithms (AofA), Krak\'ow, Poland, July 4–8, 2016,
arXiv preprint arXiv:1605.00137.


\bibitem{bitcoin}
S.~Nakamoto, \emph{Bitcoin: a peer-to-peer electronic cash system,} 2008, Available:
  \url{http://www.bitcoin.org}
  
   
\bibitem{Dwyer}
  K.J.~Dwyer and D.~Malone, \emph{Bitcoin mining and its energy footprint}. In
proceedings of the 25th IET Irish Signals Syst. Conf. (ISSC 14), Jun. 2014, pp.~280-285.



\bibitem{Pass17}
Rafael Pass, Lior Seeman, and Abhi Shelat,
\emph{Analysis of the blockchain protocol in asynchronous networks},
in proceedings of the 36th Annual International Conference on the Theory and Applications of Cryptographic Techniques (EUROCRYPT) LNCS 10211, pp.~643– 673, 2017.

\bibitem{Prodinger}
Helmut Prodinger and Guy Louchard,
\emph{The Asymmetric Leader Election Algorithm with swedish stopping: A probabilistic analysis}, Discrete Mathematics \& Theoretical Computer Science 14(2), pp.~91-128, 2012.



\bibitem{Shoker2017}
Ali Shoker, \emph{Sustainable blockchain through proof of exercise}, in proceedings of the 16th IEEE International Symposium on Network Computing and Applications (NCA), Cambridge, 2017, pp.~393-401.

  
\bibitem{ethereum}
Gavin~Wood, \emph{Ethereum: A secure decentralised generalised transaction ledger}, 2015, yellow paper.

\bibitem{mirror} K. Kelly "AR will spark the next big tech platform -- call it Mirrorworld", WIRED, 2019, https://www.wired.com/story/mirrorworld-ar-next-big-tech-platform/

\bibitem{kernel} Lundbæk, Leif-Nissen, et al. "Practical Proof of Kernel Work \& Distributed Adaptiveness." manuscript Version 1 (2018). https://pdfs.semanticscholar.org/2f1e/
cdb469f85404e141917210b2bcb6c243805a.pdf

\end{thebibliography}


\end{document}